%% file: main.tex
\documentclass{article}
\usepackage[nonatbib,final]{nips_2017}
\usepackage{blindtext, graphicx}
\usepackage[caption=false]{subfig}
\usepackage{booktabs,makecell}
\usepackage{amsmath}
\usepackage[capitalise]{cleveref}%,noabbrev
\usepackage{color}
\usepackage{cite}
\usepackage[binary-units]{siunitx}
\usepackage{threeparttable}
\usepackage{multirow}
\usepackage{tikz}
\usepackage{tikzscale}
\usepackage{rotating}
\usetikzlibrary{shapes.misc}
\usetikzlibrary{arrows}

\newcommand{\chipmunk}{\textsc{Chipmunk}}
\newcommand{\etal}{\emph{et al.}}
\newcommand{\na}{--}
\newcommand{\x}{$\times$}

\begin{document}

\title{\textsc{Chipmunk}: A Systolically Scalable 0.9\,mm${}^2$, 3.08\,Gop/s/mW\,@\,1.2\,mW Accelerator for Near-Sensor Recurrent Neural Network Inference}
\author{
    Francesco Conti\footnotemark[1]\;\footnotemark[2]\\
    \texttt{fconti@iis.ee.ethz.ch}
    \And
    Lukas Cavigelli\footnotemark[1]\\
    \texttt{cavigelli@iis.ee.ethz.ch}
    \And
    Gianna Paulin\footnotemark[1]\\
    \texttt{pauling@student.ethz.ch}
    \And
	Igor Susmelj\footnotemark[1]\\
    \texttt{isusmelj@student.ethz.ch}
	\And
	Luca Benini\footnotemark[1]\;\footnotemark[2]\\
    \texttt{lbenini@iis.ee.ethz.ch}
}
\footnotetext[1]{Integrated Systems Laboratory, ETH Zurich.}
\footnotetext[2]{Energy-Efficient Embedded Systems Laboratory, University of Bologna.}

% make the title area
\maketitle
% \bstctlcite{IEEEexam/ple:BSTcontrol}
% \renewcommand{\baselinestretch}{0.9}

\begin{abstract}
Recurrent neural networks (RNNs) are state-of-the-art in voice awareness/understanding and speech recognition.
On-device computation of RNNs on low-power mobile and wearable devices would be key to applications such as zero-latency voice-based human-machine interfaces.
Here we present \textsc{Chipmunk}, a small ($<$1\,mm${}^2$) hardware accelerator for Long-Short Term Memory RNNs in UMC 65\,nm technology capable to operate at a measured peak efficiency up to 3.08\,Gop/s/mW at 1.24\,mW peak power.
To implement big RNN models without incurring in huge memory transfer overhead, multiple \textsc{Chipmunk} engines can cooperate to form a single systolic array.
In this way, the \textsc{Chipmunk} architecture in a 75~tiles configuration can achieve real-time phoneme extraction on a demanding RNN topology proposed in \cite{Graves2013}, consuming less than 13\,mW of average power.
\end{abstract}

% \IEEEpeerreviewmaketitle

\section{Introduction}
\label{sec:intro}

In the last few years, we have witnessed an ``artificial intelligence'' revolution that has been fueled by the concurrent availability of huge amounts of training data, computing power to learn upon it, and evolution of ``smart'' algorithms, in particular those based on deep learning.
Within this field, \textit{recurrent neural networks} (RNNs), particularly Long Short-Term Memory (LSTM) and Gated Recurrent Units (GRU), are receiving increasing attention: They have shown state-of-the-art  accuracy in tasks such as speech recognition \cite{Xiong2017,Graves2013} and language translation \cite{Cho2014}, making them the forefront of the ``intelligent'' user interfaces of products such as Amazon Alexa, Google Assistant, Apple Siri, Microsoft Cortana and others.

One of the key limitations of the current generation of commercial products based on RNNs is that these embedded, edge devices depend on remote servers taking care of the computational workload necessary for the deployment of these algorithms.
Moreover, when RNNs are used as a component of human-machine interfaces, the intrinsic latency of network communication can also be problematic, as people expect the ``smart'' devices to reply not only accurately, but also timely.

For these reasons, it is very attractive to integrate RNN capabilities locally in embedded mobile and wearable platforms, making them capable of state-of-the-art voice and speech recognition autonomously and independent from external servers.
Nonetheless, while much attention has recently been dedicated to the deployment of embedded low-power inference accelerators for forward-only deep networks deployment \cite{Cavigelli2016,Conti2017,Andri2016a,Chen2016,Du2015}%[Origami, Fulmine, YodaNN, Eyeriss, ShiDianNao, ...]
, making RNNs energy-efficient is a fundamentally harder problem: the necessity to keep and update an internal state and the widespread usage of densely connected layers translate to very large memory footprint and high bandwidth requirements. 

In this work, we present a twofold contribution towards the deployment of RNN-based algorithms in devices such as smartphones, smartwatches and wearables.
First, we designed \chipmunk{}, a small and low-energy hardware accelerator engine targeted at real-time speech recognition and capable to operate autonomously on moderate size LSTM networks.
We present silicon results from a prototype chip containing a \chipmunk{} engine, which has been fabricated in UMC 65\,nm technology; the chip can achieve up to 3.8\,Gop/s at maximum efficiency operating point (@\SI{0.75}{\volt}), consuming only \SI{1.24}{\milli\watt}.

Second, we conceived a scalable computing architecture, apt to operate on bigger LSTM models as well.
As the main limitation to the deployment of big RNNs in embedded scenarios stems from their memory boundedness, we designed the \chipmunk{} engines so that they can be replicated in a systolic array, cooperating on a single bigger LSTM network.
This methodology allows the acceleration of large-scale RNNs, which can be made fast enough to operate in real-time under realistically tight time, memory and battery constraints without requiring complex, power hungry and expensive high-bandwidth main memory interfaces.

\section{Related Work}
\label{sec:related}
A recent thorough survey of efforts on hardware acceleration and design of efficient shows that few efforts have been focused on RNN inference~\cite{Sze2017}. We thus focus on this application, surveying state-of-the-art implementations from data-center to ultra-low power accelerators in the remainder of this section.

Data center workloads for RNNs are often offloaded to GPUs or specialized semi-independent co-processors such as Google's Tensor Processing Unit (TPU)~\cite{Jouppi2017} consuming in the order of 50-300\,W. The TPU is a unified architecture to target DNNs with convolutional and densely connected layers as well as LSTMs. 
However, TPUs suffer from low utilization when running RNNs. Yet 29\% of the workload running on Google's TPUs is devoted to RNN inference~\cite{Jouppi2017}, showing their relevance in commercial applications. 

In a lower power range (tens of Watts), several FPGA implementations can be found. The Efficient Speech-recognition Engine (ESE) \cite{Han2016b} targets the deployment of RNNs on a Xilinx UltraScale FPGA. To maximize efficiency and address the memory boundedness of RNNs, it heavily focuses on network quantization and pruning of the recurrent topologies and thus this accelerator engine is mainly targeted at sparse matrix-vector operations. Rybalkin~\etal~\cite{Rybalkin2017} also target bidirectional LSTMs in their FPGA accelerator. Bidirectional LSTMs have been shown to obtain better accuracy in some cases~\cite{Graves2013} but are less attractive for an online, real-time scenario as they inherently increase the network latency. Finally, DeepStream~\cite{Chang2017} is a small hardware accelerator deployed on a Xilinx Zynq 7020 targeted at text recognition with RNNs. It requires to continuously stream in weights, which makes it impractical for big RNN topologies with millions of weights.

The only published ultra-low power (few mW) implementation, the DNPU~\cite{Shin2017a}, uses two separate special-purpose engines for convolutional layers (called CP), on one side, and fully-connected and recurrent ones on the other (called FRP). The FRP does not include any particular facilities to address the stateful nature of RNNs, and it includes only a small amount of memory (\SI{10}{kB}) making external memory accesses necessary for even small RNNs, thus limiting peak performance by introducing a serious bandwidth bottleneck.

\section{Architecture}
\label{sec:archi}

\subsection{Operating principle}
\label{sec:principle}

\textit{Long Short-Term Memory} (LSTM) network layers \cite{Hochreiter1997} are often described with the following set of canonical equations:
\begin{align}
    \mathbf{i}_{t} &= \sigma( \mathbf{W}_{xi}  \mathbf{x}_t +
                              \mathbf{W}_{hi}  \mathbf{h}_{t-1} +
                              \mathbf{w}_{ci} \odot \mathbf{c}_{t-1} +
                              \mathbf{b}_{i}
                            ) \label{eq:1} \\
    \mathbf{f}_{t} &= \sigma( \mathbf{W}_{xf}  \mathbf{x}_{t} +
                              \mathbf{W}_{hf}  \mathbf{h}_{t-1} +
                              \mathbf{w}_{cf} \odot \mathbf{c}_{t-1} +
                              \mathbf{b}_{f}
                            ) \label{eq:2} \\
    \mathbf{c}_{t} &= \mathbf{f}_{t} \odot \mathbf{c}_{t-1} +
                      \mathbf{i}_{t} \odot \tanh(
                        \mathbf{W}_{xc}  \mathbf{x}_{t} +
                        \mathbf{W}_{hc} \mathbf{h}_{t-1} +
                        \mathbf{b}_{c}
                       ) \label{eq:3}  \\
    \mathbf{o}_{t} &= \sigma( \mathbf{W}_{xo}\mathbf{x}_{t} +
                              \mathbf{W}_{ho}\mathbf{h}_{t-1} +
                              \mathbf{w}_{co} \odot \mathbf{c}_{t} +
                              \mathbf{b}_{o}
                            ) \label{eq:4} \\
    \mathbf{h}_{t} &= \mathbf{o}_{t} \odot \tanh(\mathbf{c}_{t}) \label{eq:5}
\end{align}
\begin{figure}
%\centering
\resizebox{\linewidth}{!}{
\input{img/lstmDdg}
}
\caption{Data dependency graph of a LSTM. The majority of computations are the vector-matrix mult. (green) and can be distributed across multiple chips. Top-right: distribution of a vector matrix mult. to a systolic array of chips.}
\end{figure}
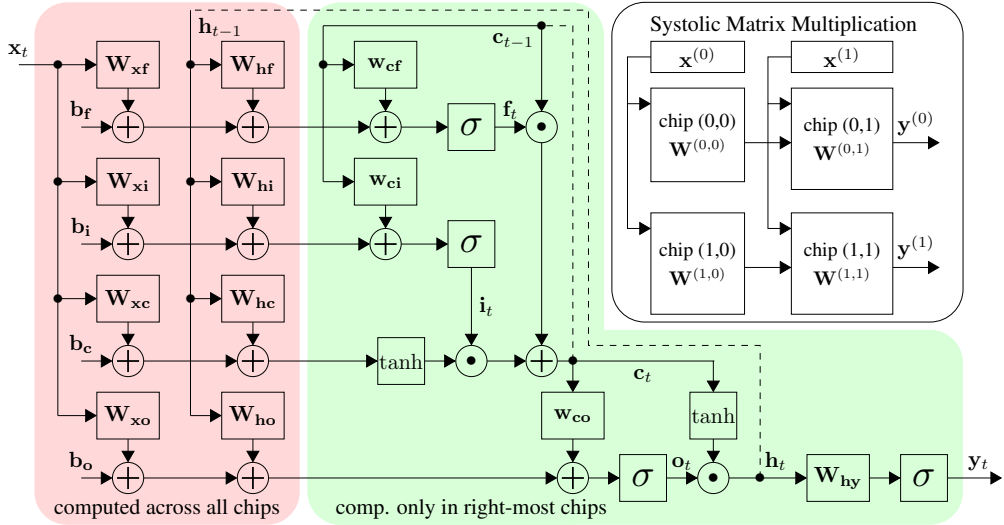
where $\mathbf{x}$ is the \textit{input state} vector;  $\mathbf{i}$, $\mathbf{f}$, $\mathbf{o}$ are called \textit{input}, \textit{forget} and \textit{output gates} respectively; $\mathbf{c}$ and $\mathbf{h}$ are the \textit{cell} and \textit{hidden states}.
The subscript indicates either the current state $t$ or the previous $t-1$, and $\odot$ denotes element-wise multiplication\footnote{
    In most literature \cref{eq:1,eq:2,eq:4} use matrix notation for $\mathbf{W}_{ci}$, $\mathbf{W}_{cf}$ and $\mathbf{W}_{co}$; however as these matrices are diagonal by construction, we use the element-wise product notation here for consistence with what is actually implemented in the \chipmunk{} hardware.
}.
The characteristic dimensions of all vectors and matrices depend on the size of the input state ($N_x$) and on that of the hidden state ($N_h$).
Multiple LSTM layers can be connected by using the hidden state of one layer as input of the next.
Finally, LSTM networks often include a final densely connected layer without recurrence: $\mathbf{y}_{t} = \sigma(\mathbf{W}_{hy} \mathbf{h}_{t})$.

In \chipmunk{}, we exploit two distinct observations regarding LSTMs. First, all compute steps are based on the same set of basic operations: \textit{i)} matrix-vector products, \textit{ii)} element-wise vector products, and \textit{iii)} element-wise non-linear activations. The internal datapath of \chipmunk{} can be configured to execute these three basic operations (\cref{sec:single_chip}) and the LSTM state parameters are stored on-chip.
Second, the vast amount of data required to compute one time step of a RNN are the weights. Storing them on-chip is thus essential to achieve high energy efficiency. To this end, we a large share of the overall chip area is dedicated to SRAM to keep the weights local. For larger LSTMs not fitting on a single chip, we allow operation in a systolic mode where the weights are split across multiple connected chips and only the much smaller intermediate results are exchanged as further discussed in \cref{sec:systolic_array}. 

\subsection{Tile architecture}
\label{sec:single_chip}

\begin{figure}[tb]
  \centering
  \subfloat[LSTM datapath.\label{fig:datapath}] {
    \centering
    \includegraphics[clip, trim= 0pt 0pt 0pt 0pt, width=0.9\textwidth] {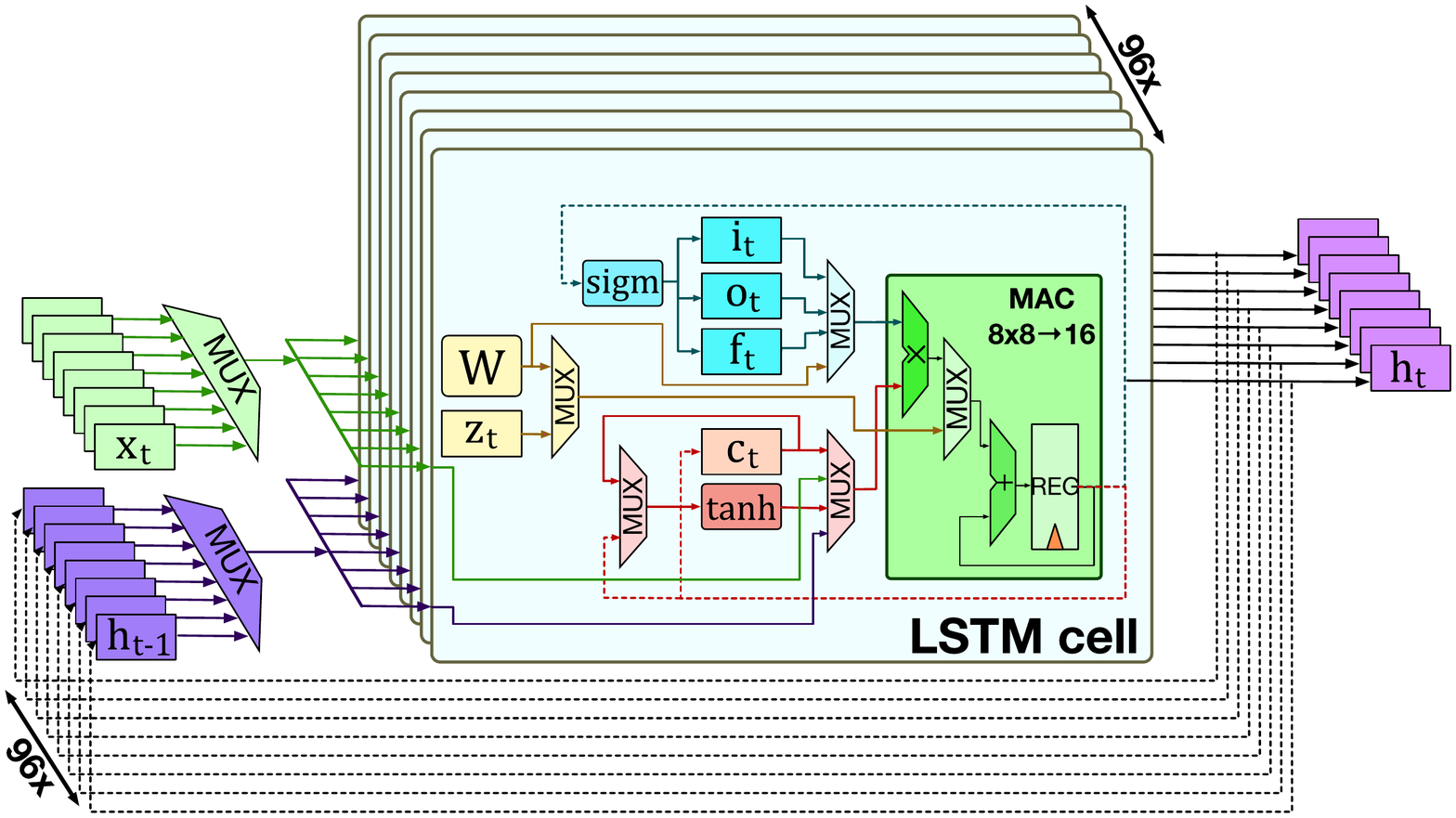}
  }
  \\
  \subfloat[Sequence of datapath basic operation loops.\label{fig:timeline}] {
    \centering
    \includegraphics[clip, trim= 0pt 0pt 0pt 0pt, width=0.9\textwidth] {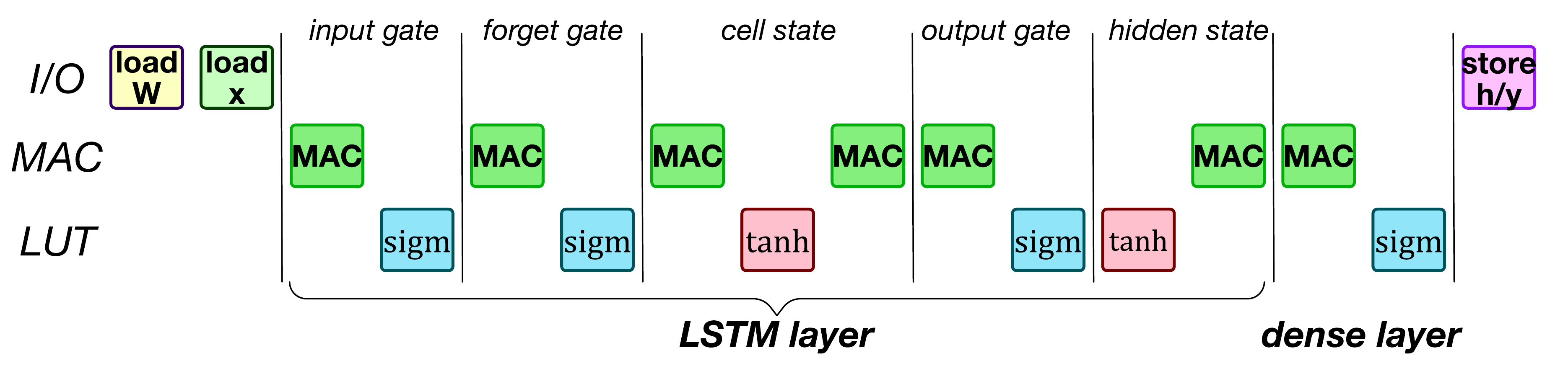}
  }
  \caption{LSTM datapath used in \chipmunk{} and typical sequence of operations. The datapath can be used to implement the operations in \cref{eq:1,eq:2,eq:3,eq:4,eq:5} by appropriately controlling the muxes and clearing the register states.}
  \label{fig:datapath_timeline}
\end{figure}

A product between a matrix of size $A\times B$ and a vector of size $B$ is composed of two nested loops, i.e. in pseudo-code:
\begin{verbatim}
    for a in range(0, A):   # row loop
      for b in range(0, B): # column loop
        z += W[a,b] * x[b]
\end{verbatim}
In \chipmunk{}, the row loop is executed on multiple parallel units, while the inner loop is executed sequentially.

\cref{fig:datapath} shows a high-level diagram of the \chipmunk{} LSTM datapath that implements this functionality.
$N_{lstm}$ parallel LSTM units are used to execute all the iterations of the row loop at the same time.
Each LSTM unit is composed of an embedded memory bank to store weights ($W$), registers for storing the $\mathbf{o}_t$, $\mathbf{f}_t$, $\mathbf{i}_t$ and $\mathbf{c}_t$ values locally, a multiply-accumulate unit and two lookup tables to implement the non-linear activation functions.
$\mathbf{x}_t$ and $\mathbf{h}_t$ are kept outside of the LSTM units, in a bank of $N_{lstm}$ registers.
At each cycle of a column loop, one element of the input state and one of the hidden state are selected depending on the iteration index and broadcast to all LSTM units.
\cref{fig:timeline} shows the basic operation loops composing a LSTM network deployed on \chipmunk{}.

All state variables use 8\,bit fixed point precision, while 16 bits are used within the multiply-accumulate block to minimize overflows.
I/O is performed via an input stream port and an output stream port, each consisting of 8\,bits of data and 2\,bits to enable a simple ready/valid handshake.
Weights are loaded at the beginning of the computation of  a LSTM layer, and inputs are streamed in sequentially.
The internal state of the LSTM cell in terms of cell state and hidden state is retained between consecutive LSTM input ``frames'' to implement the recurrent nature of the network.
A \chipmunk{} engine can be used to implement a full LSTM network with $N_x,N_h \leq N_{lstm}$ storing the weights on chip. Larger networks require to stream them in from an external source.

\subsection{Systolic scaling}
\label{sec:systolic_array}

\begin{figure}[tb]
  \centering
  \includegraphics[clip, trim= 0pt 0pt 0pt 0pt, width=0.9\textwidth] {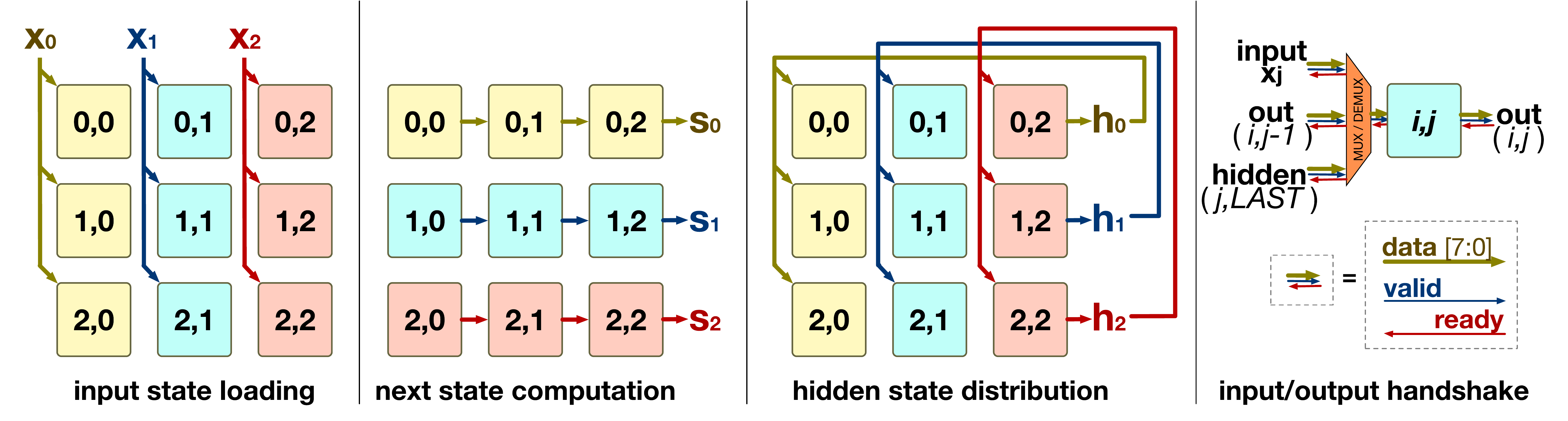}
  \caption{\chipmunk{} tile I/O and operation of a $3\times 3$ systolic array during the load of the input state $\mathbf{x}_t$, computation of the new $\mathbf{i}_{t}$, $\mathbf{o}_{t}$, $\mathbf{f}_{t}$, $\mathbf{c}_{t}$, $\mathbf{h}_{t}$ state values, and redistribution of the updated hidden state $\mathbf{h}_t$.}
  \label{fig:systolic}
\end{figure}

As the main target of the \chipmunk{} accelerator is to enable ultra-low latency applications such as on-device real-time speech recognition, the computing power of a single engine might not be sufficient.
A single engine cannot be arbitrarily scaled up: LSTM units are all coupled to the same set of registers via simple multiplexers, making it impractical to increase $N_{lstm}$ above a few hundred units.
Instead, to provide a more scalable and elegant solution, we designed \chipmunk{} so that multiple engines can be connected as tiles and share the burden of the RNN computation in a spatial fashion.

\cref{fig:systolic} shows how the computation is split between multiple tiles in the case of a $3\times 3$ array.
The input state is split into vectors of size $N_{lstm}$ and each vector is broadcast vertically along a column.
The new value for the internal gates/states is computed by accumulating the results computed by each row.
Finally, the last column can compute the output hidden state, which is broadcasted vertically to the columns for the next iteration (cf. \cref{fig:systolic}c). For a given network size/systolic configuration, these connections can be hard-wired such that no external multiplexing is required. 

\section{Results \& Discussion}
\label{sec:results}

\subsection{Silicon prototype \& Comparison with State-of-the-Art}
\label{sec:chip_results}

\begin{figure*}[tb]
    \centering
    \includegraphics[trim= 0pt 0pt 0pt 0pt, width=0.65\textwidth]{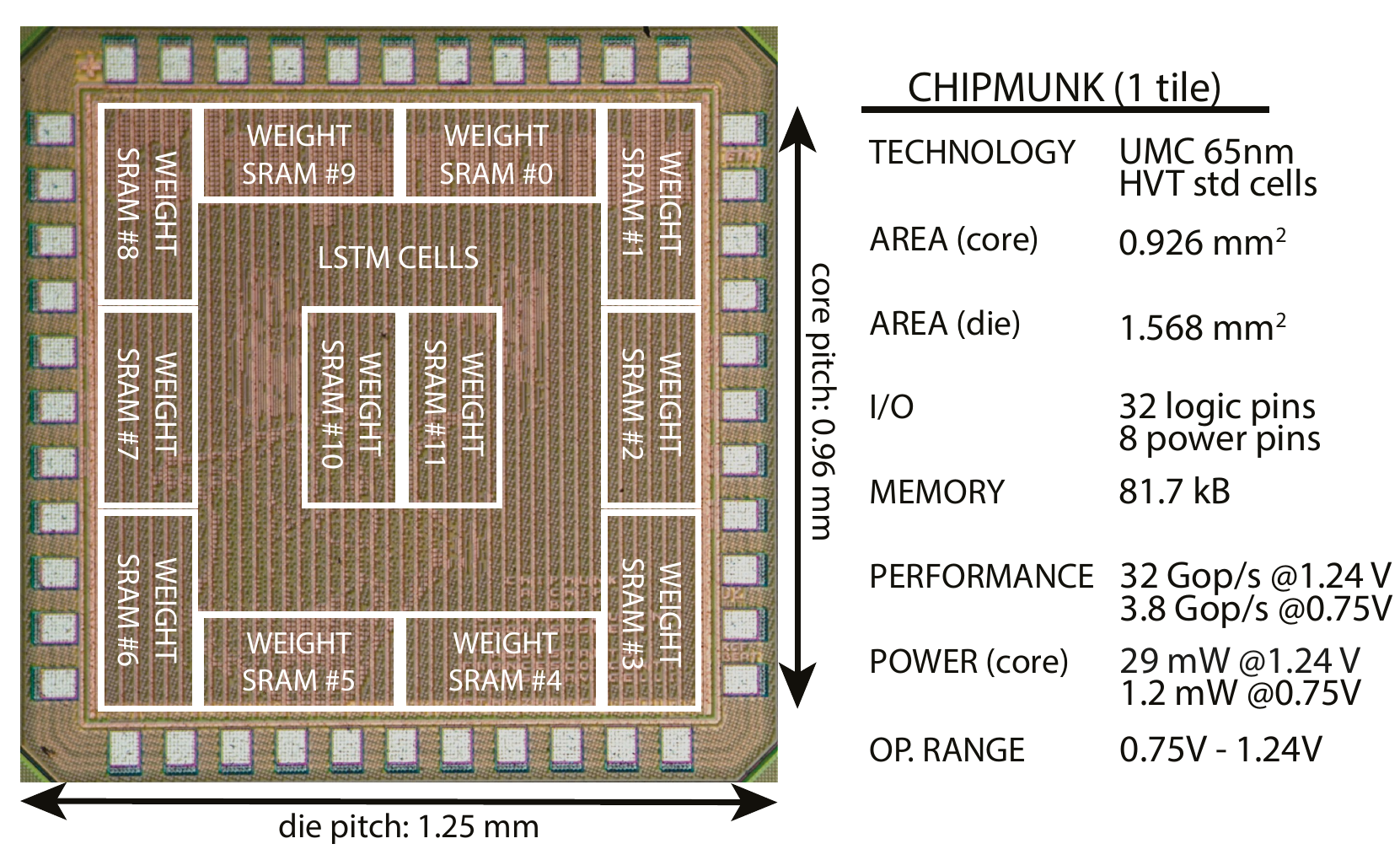}
    \caption{Microphotograph of a \chipmunk{} die.}
    \label{fig:micrograph}
\end{figure*}

\begin{figure*}[tb]
    \centering
    \includegraphics[trim= 0pt 0pt 0pt 0pt, width=0.98\textwidth]{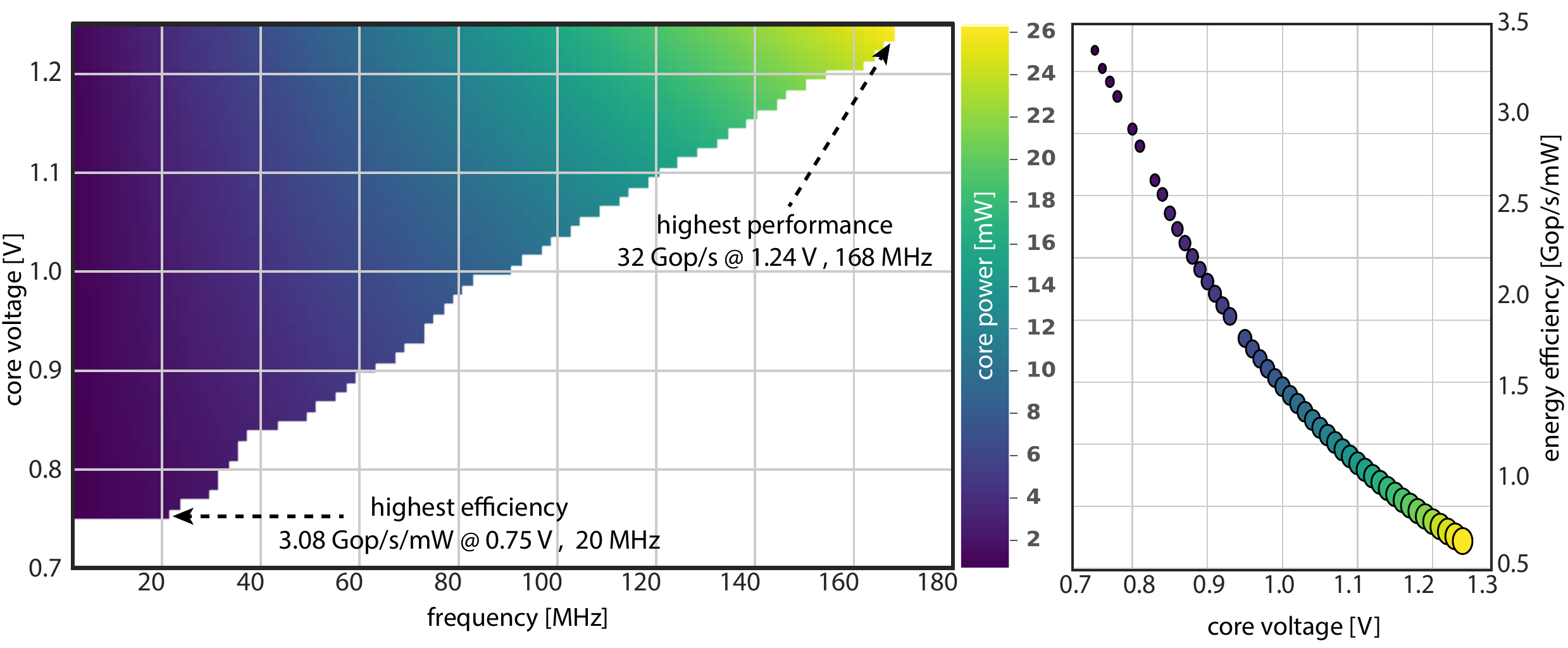}
    \caption{Frequency, power and performance of the \chipmunk{} prototype versus operating voltage at room temperature (\SI{25}{\celsius}). The left shmoo plot shows core voltage versus operating frequency; the color shade corresponds to the core power consumption (darker=less power). The right plot shows energy efficiency versus core voltage; the color shade of the scattered dots corresponds to the core power, while their size is proportional to the maximum frequency.}
    \label{fig:power_shmoo}
\end{figure*}

\input{img/hwCompTable}

We designed and built a silicon prototype based on a single \chipmunk{} tile as described in Section \ref{sec:single_chip}.
The prototype chip was fabricated in UMC 65\,nm technology, using high voltage threshold cells to minimize  leakage.
It features $N_{lstm} = 96$ LSTM units, which hold their weight and bias parameters in 12 separate SRAM banks (\SI{81.7}{\kilo\byte} in total).
The full chip, shown in  \cref{fig:micrograph}, occupies \SI{1.57}{\milli\meter^2} including the pads.
The chip exposes the interface described in Section \ref{sec:systolic_array} for tile-to-tile communication, so that it would be possible to prototype a systolic array using many discrete chips.

\cref{fig:power_shmoo} shows the experimental results obtained by testing the \chipmunk{} prototype at room temperature (\SI{25}{\celsius}).
The prototype is fully functional in an operating range between \SI{0.75}{\volt} (limited by SRAM failure) and \SI{1.24}{\volt}, corresponding to a range of \num{20} to \SI{168}{MHz} of maximum clock frequency and from \num{1.24} to \SI{29}{mW} of power consumption.
The peak performance in terms of operations per second\footnote{As customary for neural network accelerators, we count 1 multiply-accumulate as 2 operations.} of one \chipmunk{} chip is 32.2\,Gop/s (at \SI{1.24}{\volt}) and the peak energy efficiency (3.08\,Gop/s/mW) is reached at \SI{0.75}{\volt}.

\cref{tbl:hwComp} compares architectural parameters and synthetic results between \chipmunk{} and the existing VLSI and FPGA-based implementations for which performance and energy numbers have been published.
Our work reaches comparable performance with the DNPU proposed by Shin~\etal~\cite{Shin2017a}.
Performance is obviously below that claimed by Google TPU~\cite{Jouppi2017}, but this is mostly due to the different size.
In fact, despite the TPU uses 28\,nm integration, \chipmunk{} has 2.8$\times$ better area efficiency - and a performance-wise ``TPU-equivalent'' array with $\sim$115 \chipmunk{} engines would consume only \SI{3.33}{\watt}, an order of magnitude less than the TPU.
\chipmunk{} advances the state-of-the-art energy efficiency with respect to the DNPU, showing a 39\% improvement. Moreover, the DNPU does not include any provision to address the fundamental memory boundedness of RNNs, which  \chipmunk{} addresses via systolic scaling.
All FPGA implementations \cite{Han2016b,Rybalkin2017,Chang2017} are at least two orders of magnitude less energy-efficient.

In terms of arithmetic precision we have chosen to use 8\,bit fixed-point representations for storage and perform the MAC operations with 16\,bit precision. This is in line with Google's TPU and higher than the 4-7\,bit of the DNPU. 

\subsection{Real-world speech recognition}
\label{sec:speech}

\begin{table}
    \centering
    \caption{CTC-3L-421H-UNI Speech Recognition LSTM Executed on \chipmunk{} With a 10\,ms Constraint}
\label{tbl:systolic}
\begin{tabular}{l|rrr}
\toprule
                                         & Configuration  & PERF @\SI{1.24}{V} & EFF @\SI{0.75}{V} \\ \midrule
\multirow{3}{*}{\textit{Execution time}} & systolic 3\x 5\x 5          & \textbf{0.09~ms} & \textbf{0.76~ms} \\
                                         & systolic 5\x 5          & \textbf{1.59~ms} & 13.31~ms \\
                                         & single          & 38.23~ms & 321.14~ms \\ \midrule
\multirow{3}{*}{\textit{Peak power}}     & systolic 3\x 5\x 5          & \textbf{1833.75~mW} & \textbf{165.75~mW} \\
                                         & systolic 5\x 5          & \textbf{611.25~mW} & 55.25~mW \\
                                         & single          & 24.45~mW & 2.21~mW \\ \midrule
\multirow{3}{*}{\textit{Average power}}  & systolic 3\x 5\x 5          & \textbf{16.53~mW} & \textbf{12.55~mW} \\
                                         & systolic 5\x 5          & \textbf{96.89~mW} & - \\
%                                         & single          & - & - \\ 
\bottomrule
\end{tabular}
\end{table}

To evaluate \chipmunk{} on a real-world problem, we targeted \emph{CTC-3L-421H-UNI}, a 3-layer, 421-hidden units per layer LSTM topology introduced by Graves \etal \cite{Graves2013}, which takes as input a stream of 123 Mel-Frequency Cepstral Coefficients (MFCCs) extracted from an audio stream and identifies phonemes with an error rate of 19.6\%, evaluated on the TIMIT database.
The MFCC input ``frames'' are produced with a \SI{10}{\milli\second} rate, which means that any embedded low-latency real-time RNN implementation should be able to elaborate the full network in less than this time.
We evaluate three different \chipmunk{} configurations:  a systolic array of 75 units, divided in 3 sub-arrays of $5\times 5$ engines; a single array of $5\times 5$ engines; and a single \chipmunk{} engine.
The largest configuration can host the full topology in a spatial fashion; each of the sub-arrays hosts one layer of the RNN.
After the initial programming phase, it does not need any reprogramming.
The smaller arrays need to be reconfigured at each new layer (in the $5\times 5$ array case) or multiple times per layer (in the single unit case).

\cref{tbl:systolic} reports execution time and power for these three configurations.
Execution times include both computation and reconfiguration, excluding only the initial configuration which doesn't need to be repeated for each new frame/layer.
Bold time/power values indicate configurations that can meet the \SI{10}{\milli\second} deadline.
As the \textit{CTC-3L-421H-UNI} topology has $\sim 3.8\times 10^6$ weights, a $3\times 5\times 5$ systolic configuration is best used (all weights stored locally). Smaller configurations imply a $>80\%$ overhead for reloading weights.

Average power, also shown in \cref{tbl:systolic}, is computed under the assumption that the array is perfectly duty cycled when not in use over the \SI{10}{ms} window.
Even in the assumption that the \chipmunk{} array is always-on, the 12.55\,mW required to process this network would only add $\sim$4\% to idle power on a typical smartphone (300 to 400\,mW~\cite{Carroll2010}).
Adding a filter to drop clearly uninteresting input (e.g. silence) would likely decrease this overhead by an order of magnitude.

\section{Conclusion}
We have presented an architecture and silicon measurement results for a small (0.9\,mm${}^2$) RNN hardware accelerator providing 3.8\,Gop/s at 1.2\,mW in 65\,nm digital CMOS technology, resulting in new state-of-the-art energy and area efficiencies of 3.08\,Gop/s/mW and 34.4\,Gop/s/mm${}^2$.
The systolic design is scalable to accommodate also large RNNs efficiently by connecting multiple identical chips on the circuit board.

\section*{Acknowledgements}
This work was supported in part by the EU project ExaNoDe under grant H2020-671578, and in part by the Swiss National Science Foundation project Micropower Deep Learning. 

\bibliographystyle{IEEEtran}
\bibliography{bstctl,IEEEabrv,mendeley}

\end{document}

%% file: img/lstmDdg.tex
\usetikzlibrary{shapes.misc}
\usetikzlibrary{arrows}
\tikzset{
  circ/.style={
    cross out, 
    minimum size=0.25cm, 
    inner sep=0pt, 
    outer sep=0pt, 
    thick, draw=black, 
    align=center
  }
}
\tikzset{
  pics/circadd/.style={
    code={
    	\node[circ, rotate=45]{};
    	\draw (0,0) circle (0.25);
    }
  }
}
\tikzset{
  pics/circmul/.style={
    code={
    	\node[]{$\bullet$};
    	\draw (0,0) circle (0.25);
    }
  }
}
\tikzset{
  pics/const/.style={
    code={
    	\node at (-0.65,0) {$\mathbf{b_{#1}}$};
    	\draw (-1,-0.5) -- (-1,0.5) -- (0,0) -- cycle;
    }
  }
}

%\tikzstyle{circ} = [cross out, minimum size=0.25cm, inner sep=0pt, outer sep=0pt, thick, draw=black, align=center]
% \tikzstyle{circ2} = [code={
% 		\node [circ] at (0,0) {};
% 	}] 
\begin{tikzpicture}

%\clip (15.75,-8.5) rectangle (-0.375,0.125);

\draw [fill=red!15!white, draw=none, rounded corners=15pt] (0.125,0) rectangle (4.375,-8.375);
\node at (2.25,-8.125) {computed across all chips};
\draw [fill=green!15!white, draw=none, rounded corners=15pt] (4.5,0) rectangle (9.25,-8.375);
\draw [fill=green!15!white, draw=none, rounded corners=15pt] (4.5,-5.25) rectangle (15,-8.375);
\node at (7.125,-8.125) {comp. only in right-most chips};
\draw [draw=black, rounded corners=15pt] (9.375,0) rectangle (15,-5.125);
\node at (12.125,-0.375) {Systolic Matrix Multiplication};

% 	\pgfmathsetmacro{\ygrid}{2.5}
% 
% 	\foreach \x in {0,...,2} {
% 		
% 	`W\draw [red!60!black] (3*\x-0.75,3.0) node [above] {$\mathbf{x}^{(\x)}$} -- (3*\x-0.75,-4.5);
% 	    
% 	    \foreach \y in {0,...,2} {
% 			%\pgfmathtruncatemacro{\label}{\x - 5 *  \y +21}
%        		%\node []  (\x\y) at (1.5*\x,1.5*\y) {\label};
%        		\draw  (3*\x,-\ygrid*\y) rectangle (3*\x+2,-\ygrid*\y+2) node [pos=0.5]{$\mathbf{W^{(\y,\x)}}$};
% 			%\pgfmathtruncatemacro{\label}{\x - 5 *  \y +21}
%        		%\node []  (\x\y) at (1.5*\x,1.5*\y) {\label};
%        		\draw [red!60!black,-triangle 60] (3*\x-0.75,-\ygrid*\y+0.5) -- (3*\x,-\ygrid*\y+0.5);
%        		\draw [-triangle 60] (3*\x-0.5,-\ygrid*\y+1) -- (3*\x,-\ygrid*\y+1);
%        		\ifthenelse{\x<2}{
%        			\draw [green,-triangle 60] (3*\x+2,-\ygrid*\y+1.5) -- (3*\x+3,-\ygrid*\y+1.5);
%        		}{}
%        	} 
%        }
%        	
% 	\foreach \y in {0,...,2} {
% 		\draw (8,1-\ygrid*\y) node [xshift=10,yshift=-7] {$\mathbf{h}^{(\y)}$} -- (8.5+0.25*\y,1-\ygrid*\y) -- (8.5+0.25*\y,2.75-0.25*\y) -- (-0.5+3*\y,2.75-0.25*\y) -- (-0.5+3*\y,-4);
% 		%\draw (8,1-\ygrid*\y) -- (8.5,1-\ygrid*\y) -- (8.5,2.5) -- (2.75,2.5) -- (2.75,-5.0);
% 		%\draw (8,1-\ygrid*\y) -- (8.75,1-\ygrid*\y) -- (8.75,2.25) -- (5.75,2.25) -- (5.75,-5.0);
%        }

\draw (1.125,-0.625) rectangle (2.125,-1.375) node [pos=0.5] {$\mathbf{W}_\mathbf{xf}$};
\draw (3.125,-0.625) rectangle (4.125,-1.375) node [pos=0.5] {$\mathbf{W}_\mathbf{hf}$};
\draw (5.25,-0.625) rectangle (6.25,-1.375) node [pos=0.5] {$\mathbf{w}_\mathbf{cf}$};
\draw (6.75,-1.625) rectangle (7.5,-2.375) node [pos=0.5, scale=1.76] {$\sigma$};
\draw (1.625,-2) pic{circadd} {};
\draw (5.75,-2) pic{circadd} {};
\draw (3.625,-2) pic{circadd} {};
\draw (8.25,-2) pic{circmul} {};
% \draw (-0.125,-1.375) node (v1) {} pic{const={f}} {};
\draw [-triangle 60] (0.875,-2) node [above] {$\mathbf{b_f}$} -- (1.375,-2);
\draw [-triangle 60] (1.875,-2) -- (3.375,-2);
\draw [-triangle 60] (3.875,-2) -- (5.5,-2);
\draw [-triangle 60] (6,-2) -- (6.75,-2);
\draw [-triangle 60] (7.5,-2) -- (8,-2) node[midway, above]{$\mathbf{f}_t$};
\draw [-triangle 60] (1.625,-1.375) -- (1.625,-1.75);
\draw [-triangle 60] (3.625,-1.375) -- (3.625,-1.75);
\draw [-triangle 60] (5.75,-1.375) -- (5.75,-1.75);
\draw [-triangle 60] (4.75,-0.375) --(8.25,-0.375) -- (8.25,-1.75);
\node at (2.625,-1) {$\bullet$};
\node at (4.75,-1) {$\bullet$};
\node at (0.5,-1) {$\bullet$};
\node at (8.25,-0.375) {$\bullet$};
\draw [-triangle 60](0.5,-1) -- (1.125,-1);
\draw [-triangle 60](2.625,-1) -- (3.125,-1);
\draw [-triangle 60](4.75,-1) -- (4.75,-1) -- (5.25,-1);

\draw (1.125,-6.25) rectangle (2.125,-7) node [pos=0.5] {$\mathbf{W}_\mathbf{xo}$};
\draw (3.125,-6.25) rectangle (4.125,-7) node [pos=0.5] {$\mathbf{W}_\mathbf{ho}$};
\draw (9.5,-7.25) rectangle (10.25,-8) node [pos=0.5, scale=1.76] {$\sigma$};
\draw (14,-7.25) rectangle (14.75,-8) node [pos=0.5, scale=1.76] {$\sigma$};
\draw (1.625,-7.625) pic{circadd} {};

\draw (3.625,-7.625) pic{circadd} {};
\draw (8.75,-7.625) pic{circadd} {};
\draw (11,-7.625) pic{circmul} {};
% \draw (-0.125,-7) node (v1) {} pic{const={o}} {};
%\node [circadd] at (3.5,-2) {};

\draw [-triangle 60] (0.875,-7.625) node [above] {$\mathbf{b_o}$} -- (1.375,-7.625);
\draw [-triangle 60] (1.875,-7.625) -- (3.375,-7.625);
\draw [-triangle 60] (3.875,-7.625) -- (8.5,-7.625);

\draw [-triangle 60] (8.5,-5.75) -- (11,-5.75) -- (11,-6.25);
\draw [] (11,-5.75) -- (11,-5.75);
\draw [-triangle 60] (8.75,-5.75) -- (8.75,-6.25);
\draw [-triangle 60] (8.75,-7) -- (8.75,-7.375);
\draw [-triangle 60] (11,-7) -- (11,-7.375);
\node at (8.75,-5.75) {$\bullet$};
% \node at (11,-5.75) {$\bullet$};
\draw [-triangle 60] (9,-7.625) -- (9.5,-7.625);
\draw [-triangle 60] (10.25,-7.625) -- (10.75,-7.625) node[midway, above]{$\mathbf{o}_t$};
\draw [-triangle 60] (13.5,-7.625) -- (14,-7.625);
\draw [-triangle 60] (1.625,-7) -- (1.625,-7.375);
\draw [-triangle 60] (3.625,-7) -- (3.625,-7.375);

\draw [](-0.125,-1) node [above] {$\mathbf{x}_t$} --(0.5,-1) -- (0.5,-6.625);
\draw [](2.625,-0.125) node [below right] {$\mathbf{h}_{t-1}$}-- (2.625,-6.625);
\draw [](4.75,-0.375)-- (4.75,-2.875);

\draw [-triangle 60](0.5,-6.625) -- (1.125,-6.625);
\draw [-triangle 60](2.625,-6.625) -- (3.125,-6.625);

\draw (1.125,-4.375) rectangle (2.125,-5.125) node [pos=0.5] {$\mathbf{W}_\mathbf{xc}$};
\draw (3.125,-4.375) rectangle (4.125,-5.125) node [pos=0.5] {$\mathbf{W}_\mathbf{hc}$};
\draw (5.625,-5.375) rectangle (6.375,-6.125) node [pos=0.5, scale=1] {$\tanh$};
\draw (10.625,-6.25) rectangle (11.375,-7) node [pos=0.5, scale=1] {$\tanh$};
\draw (1.625,-5.75) pic{circadd} {};
\draw (3.625,-5.75) pic{circadd} {};
\draw (8.25,-5.75) pic{circadd} {};
\draw (7.125,-5.75) pic{circmul} {};
% \draw (-0.125,-5.125) node (v1) {} pic{const={c}} {};
\draw [-triangle 60] (0.875,-5.75) node [above] {$\mathbf{b_c}$} -- (1.375,-5.75);
\draw [-triangle 60] (1.875,-5.75) -- (3.375,-5.75);
\draw [-triangle 60] (3.875,-5.75) -- (5.625,-5.75);
\draw [-triangle 60] (6.375,-5.75) -- (6.875,-5.75);
\draw [-triangle 60] (7.375,-5.75) -- (8,-5.75);
\draw [-triangle 60] (1.625,-5.125) -- (1.625,-5.5);
\draw [-triangle 60] (3.625,-5.125) -- (3.625,-5.5);
\node at (2.625,-4.75) {$\bullet$};
\node at (0.5,-4.75) {$\bullet$};
\draw [-triangle 60](0.5,-4.75) -- (1.125,-4.75);
\draw [-triangle 60](2.625,-4.75) -- (3.125,-4.75);
\draw (8.25,-6.25) rectangle (9.25,-7) node [pos=0.5] {$\mathbf{w}_\mathbf{co}$};

\draw (1.125,-2.5) rectangle (2.125,-3.25) node [pos=0.5] {$\mathbf{W}_\mathbf{xi}$};
\draw (3.125,-2.5) rectangle (4.125,-3.25) node [pos=0.5] {$\mathbf{W}_\mathbf{hi}$};
\draw (5.25,-2.5) rectangle (6.25,-3.25) node [pos=0.5] {$\mathbf{w}_\mathbf{ci}$};
\draw (6.75,-3.5) rectangle (7.5,-4.25) node [pos=0.5, scale=1.76] {$\sigma$};
\draw (1.625,-3.875) pic{circadd} {};
\draw (5.75,-3.875) pic{circadd} {};
\draw (3.625,-3.875) pic{circadd} {};
% \draw (-0.125,-3.25) node (v1) {} pic{const={i}} {};
\draw [-triangle 60] (0.875,-3.875) node [above] {$\mathbf{b_i}$} -- (1.375,-3.875);
\draw [-triangle 60] (1.875,-3.875) -- (3.375,-3.875);
\draw [-triangle 60] (3.875,-3.875) -- (5.5,-3.875);
\draw [-triangle 60] (6,-3.875) -- (6.75,-3.875);
\draw [-triangle 60] (7.125,-4.25) -- (7.125,-5.5) node[midway, right]{$\mathbf{i}_t$};
\draw [-triangle 60] (1.625,-3.25) -- (1.625,-3.625);
\draw [-triangle 60] (3.625,-3.25) -- (3.625,-3.625);
\draw [-triangle 60] (5.75,-3.25) -- (5.75,-3.625);
\node at (2.625,-2.875) {$\bullet$};
\node at (0.5,-2.875) {$\bullet$};
\node at (11.75,-7.625) {$\bullet$};
\draw [-triangle 60](0.5,-2.875) -- (1.125,-2.875);
\draw [-triangle 60](2.625,-2.875) -- (3.125,-2.875);
\draw [-triangle 60](4.75,-2.875) -- (4.75,-2.875) -- (5.25,-2.875);

\draw [-triangle 60] (8.25,-2.25) -- (8.25,-5.5);

\draw [-triangle 60] (11.25,-7.625) -- (12.5,-7.625);
\draw [-triangle 60] (14.75,-7.625) -- (15.625,-7.625);

\node [below] at (9.875,-5.75) {$\mathbf{c}_t$};
\node [above] at (12,-7.625) {$\mathbf{h}_t$};
\node [above] at (15.25,-7.625) {$\mathbf{y}_t$};
\draw [dashed](8.75,-5.75) -- (8.75,-5.75) -- (8.75,-5.125) -- (8.75,-0.375) -- (8.25,-0.375) node [below left] {$\mathbf{c}_{t-1}$};
\draw [dashed](11.75,-7.625) -- (11.75,-5.5) -- (9,-5.5) -- (9,-0.125) -- (2.625,-0.125);

\draw (12.5,-7.25) rectangle (13.5,-8) node [pos=0.5] {$\mathbf{W}_\mathbf{hy}$};

\draw (10,-1.375) rectangle (11.5,-2.875) node [pos=0.5, align=center, scale=0.9] {chip (0,0)\\[1mm]$\mathbf{W}^{(0,0)}$};
\draw (10,-3.375) rectangle (11.5,-5) node [pos=0.5, align=center, scale=0.9] {chip (1,0)\\[1mm]$\mathbf{W}^{(1,0)}$};
\draw (12.25,-3.375) rectangle (13.875,-5) node [pos=0.5, align=center, scale=0.9] {chip (1,1)\\[1mm]$\mathbf{W}^{(1,1)}$};
\draw (12.25,-1.375) rectangle (13.875,-3) node [pos=0.5, align=center, scale=0.9] {chip (0,1)\\[1mm]$\mathbf{W}^{(0,1)}$};

\draw (10,-0.625) rectangle (11.5,-1.125) node [pos=0.5] {$\mathbf{x}^{(0)}$};
\draw (12.25,-0.625) rectangle (13.875,-1.125) node [pos=0.5] {$\mathbf{x}^{(1)}$};

\draw [-triangle 60] (10,-0.875) -- (9.625,-0.875) -- (9.625,-1.625) -- (10,-1.625);
\draw [-triangle 60] (9.625,-1.625) -- (9.625,-3.625) -- (10,-3.625);

\draw [-triangle 60] (12.25,-0.875) -- (11.875,-0.875) -- (11.875,-1.625) -- (12.25,-1.625);
\draw [-triangle 60] (11.875,-1.625) -- (11.875,-3.625) -- (12.25,-3.625);

\draw [-triangle 60, black]  (11.5,-2.25) -- (12.25,-2.25);
\draw [-triangle 60, black]  (11.5,-4.25) -- (12.25,-4.25);
\draw [-triangle 60, black]  (13.875,-2.25) -- (14.625,-2.25) node [midway, above] {$\mathbf{y}^{(0)}$};
\draw [-triangle 60, black]  (13.875,-4.25) -- (14.625,-4.25) node [midway, above] {$\mathbf{y}^{(1)}$};
\end{tikzpicture}

%% file: img/hwCompTable.tex
\begin{sidewaystable*}
    \centering
    \caption{Comparison to Existing VLSI and FPGA Implementations}
    \label{tbl:hwComp}
    \begin{threeparttable}
        \begin{tabular}{l|rrr|rrr}
\toprule
& \textbf{THIS WORK} & DNPU\tnote{*}~\cite{Shin2017a} & Google TPU\tnote{\dag}~\cite{Jouppi2017} & Han\,\etal\tnote{\ddag}~\cite{Han2016b} & Rybalkin \etal~\cite{Rybalkin2017} & Chang \etal~\cite{Chang2017} \\ 
%& \textbf{THIS WORK} & Shin et al. (DNPU) \cite{Shin2017} & Han et al. (ESE) \cite{Han2016b} & Guan et al. (Cong) \cite{Guan2017} & Ribarkin et al. (Wehn) \cite{Rybalkin2017} & Chank and Culurciello \cite{Chang2017} \\ 
\midrule
Technology 
& UMC 65\,nm CMOS & 65\,nm CMOS & 28\,nm CMOS & Xilinx XCKU060 
& Xilinx Z7045 & Xilinx Z7020 \\
%Area (core/die) [mm2] 
%& 0.93mm2 core / 1.57mm2 die & 2.0mm2 core / 16.0mm2 die & 294k LUT, 453k FF 
%& 1176 DSP, 182k LUT, 190k FF & 33 DSP, 33k LUT, 15k FF & ? \\
Area 
& core: 0.93\,mm${}^2$ & core\tnote{*}: $\sim$2.0\,mm${}^2$ & \na & 294k\,LUT
& 33k\,LUT & 7.6k\,LUT \\
%& die: 1.57\,mm${}^2$ & die: 16.0\,mm${}^2$ & 453k\,FF 
%& 190k\,FF & 15k\,FF \\
& die: 1.57\,mm${}^2$ & die: 16.0\,mm${}^2$ & die: $\sim$300\,mm${}^2$ & 453k\,FF 
& 15k\,FF, 33\,DSP & 13k\,FF, 50\,DSP \\
%& &  & & 1176\,DSP & 33\,DSP &  \\
On-chip memory
& 82\,kB & 10\,kB & 28\,MB & 4.2\,MB & 332\,kB & \na \\
Arithmetic 
& 8-16\,bit & 4-7\,bit & 8-16\,bit & 12\,bit, pruned & 5-16\,bit & 16\,bit \\
Number of MACs & 96 & 64 & 66k & \na & \na & 4 \\
Core voltage 
& 1.24\,V / 0.75\,V & 1.1\,V / 0.77\,V & \na & \na & \na & \na \\
Frequency 
& 168 / 20\,MHz & 200 / 50\,MHz & 700\,MHz & 200\,MHz & 166\,MHz & 142\,MHz\\
Power 
& 29.03 / \textbf{1.24\,mW} & 21 / 2.6\,mW & 40-28\,W & 41\,W & $\sim$10\,W & 2.3\,W \\
Peak performance
& 32.3 / 3.8\,Gop/s & 25 / 6.25\,Gop/s & \textbf{3.7-2.8\,Top/s} & 2.5\,Top/s (equiv.)\tnote{\ddag} & 152\,Gop/s & 0.389\,Gop/s \\
Energy efficiency
& 1.11 / \textbf{3.08\,Gop/s/mW} & 1.10 / 2.22\,Gop/s/mW & $<$0.13\,Gop/s/mW  & 0.061\,Gop/s/mW\tnote{\ddag} & 0.0152\,Gop/s/mW & 0.000146\,Gop/s/mW \\
Area efficiency
& \textbf{34.4\,Gop/s/mm${}^2$} & 12.5\,Gop/s/mm${}^2$ & 12.3-9.3\,Gop/s/mm${}^2$ & \na & \na & \na \\ \bottomrule
        \end{tabular}
        \begin{tablenotes}
            \item [*] The DNPU is a mixed CNN-RNN processor. We report here only the figures related to the RNN subunit. 
            \item [\dag] We present here the values from \cite{Jouppi2017} based on the two LSTMs for which they measured the performance. For both, the TPU is severely memory bandwidth limited. 
            \item [\ddag] They assume a well-structured sparsity of 11.2\% in the weight matrices. Reported numbers are dense-equivalent throughput. Underlying compute throughput: 282\,Gop/s. 
        \end{tablenotes}
    \end{threeparttable}
\end{sidewaystable*}